\documentclass[12pt,preprint]{aastex}
\shorttitle{NIR Polarimetry of $\beta$ Pic}
\shortauthors{Tamura et al.}

\begin{document}

\title{FIRST TWO-MICRON IMAGING POLARIMETRY OF BETA PICTORIS}

\author{Motohide Tamura\altaffilmark{1, 2}, Misato Fukagawa\altaffilmark{1, 3, 4}}
\author{Hiroshi Kimura\altaffilmark{5}, Tetsuo Yamamoto\altaffilmark{5}}
\author{Hiroshi Suto\altaffilmark{1}, Lyu Abe\altaffilmark{1}}


\altaffiltext{1}{National Astronomical Observatory of Japan, Osawa, Mitaka, Tokyo 
181-8588, Japan}
\altaffiltext{2}{Department of Astronomical Science, Graduate University for 
Advanced Studies (Sokendai), Osawa, Mitaka, Tokyo 181-8588, Japan}
\altaffiltext{3}{Department of Astrophysics, Faculty of Sciences, Nagoya University, 
Chikusa-ku, Nagoya 464-8602, Japan}
\altaffiltext{4}{Spitzer Science Center, California Institute of Technology,
MC 220-6, Pasadena CA 91125}
\altaffiltext{5}{Institute of Low Temperature Science,
Hokkaido University, Sapporo 060-0819, Japan}

\begin{abstract}
High-resolution \textit{K} band imaging polarimetry of the $\beta$ Pic dust disk
has been conducted with adaptive optics and a coronagraph using the Subaru
8.2-m telescope.
Polarization of $\sim$10 \% is detected out to
$r \sim$ 120 AU with a centro-symmetric vector pattern
around the central star, confirming that the disk
is seen as an infrared reflection nebula.
We have modeled our near-infrared and previous optical polarization results
in terms of dust scattering in the disk and have found that
both the degrees of polarization
and the radial intensity profiles are well reproduced.
We argue that the observed characteristics of the disk dust are consistent with 
the presence of ice-filled
fluffy aggregates consisting of submicron grains
in the $\beta$ Pic system.
There is a gap around 100 AU in both the intensity and polarization profiles,
which suggests a paucity of planetesimals in this region.
The radial intensity profile also shows ripple-like structures, which are
indicative of the presence of multiple planetesimal belts,
as in the case of the M-type Vega-like star AU Mic.
\end{abstract}


\section{INTRODUCTION}
The A5 V star $\beta$ Pic (RA = 5$^{\mathrm h}$ 47$^{\mathrm m}$ 17\fs 08, 
DEC = {$-51$}$^\circ$ 3$'$ 59\farcs 5 (J2000.0); $V$ = 3.85, $K$ = 3.53, 
$d$ = 19.28 pc, age $\sim$20 Myr) was among the first main-sequence stars for which 
significant excess infrared emission was detected by
the {\it Infrared Astronomical Satellite} 
($IRAS$) (Aumann et al. [1984]
; see Crifo et al. [1997] for distance and Barrado y Navascu\'{e}s et al. [1999] for age). 
Subsequent observations have revealed 
that the infrared emission is due to the presence of a dusty disk around the star. 
Because the faint circumstellar disk lies near the very bright central  
star, 
various techniques for achieving a high contrast have 
been employed to directly observe the circumstellar structure. 
The first successful observations were made with an optical CCD coronagraph, 
which clearly detected scattered light from a nearly edge-on dust disk 
(Smith \& Terrile 1984), then followed by anti-blooming CCD imaging, 
new optical coronagraph imaging, 
near-infrared adaptive optics (AO) imaging,
and {\it Hubble Space Telescope} ($HST$) optical imaging 
(Lacavelier des Etangs et al. 1993; 
Golimowski et al. 1993; Kalas \& Jewitt 1995; 
Mouillet et al. 1997a, 1997b; Heap et al. 2000). 
These observations show various small scale structures in the  
scattering disk extending from $r \sim$ 15 to $\sim$1800 AU.
Recent mid-infrared and submillimeter imaging observations have also  
overcome the contrast problem by tracing the thermal emission component  
from the disk because the central star is relatively faint at longer  
wavelengths (Lagage \& Pantin 1994; Pantin et al. 1997; Wahhaj et al.  
2003; Okamoto et al. 2004; Telesco et al. 2005; Holland et al. 1998).
The surface brightness profile in the disk midplane changes abruptly with 
its radial slope at $r \sim 100$~AU.
Dust grains in the disk are thought to be generated 
from larger bodies such as planetesimals, which resemble comets and asteroids 
in our solar system.

Imaging polarimetry is one of the most useful methods for studying the structure 
and dust properties of such extended circumstellar disks 
(e.g., Artymowicz 1997). 
However, in spite of being the most studied Vega-type stars, 
$\beta$ Pic has rarely been the target of polarimetric observations. 
Gledhill et al. (1991) first performed polarimetric imaging in the \textit{R} band 
with a resolution of about 1\farcs 5. 
Wolstencroft et al. (1995) 
extended the wavelengths to \textit{B}, \textit{V}, \textit{R}, 
and \textit{I} bands. 
These optical polarization data have been modeled by 
Voshchinnikov \& Kr\"{u}gel (1999) and Krivova et al. (2000).
The model results indicate that micron-size grains are the most  
important contributor to scattering and polarization of stellar radiation.
To date, no near-infrared (either aperture or imaging) polarimetry studies
have been reported, even though the  properties of these optically 
dominant micron-size grains are best  
studied at near-infrared wavelengths.

In this paper, we present results on the AO coronagraphic imaging  
polarimetry of $\beta$ Pic in the $K$ band (2.2 $\mu {\rm m}$) obtained  
with the 8.2 m Subaru telescope and the infrared coronagraph imager 
with adaptive optics (CIAO).  
We discuss the properties of the dust in the disk on the basis of the polarization and  
intensity data by employing a light-scattering dust model.
Although the central $\sim$1$''$ region is still unobservable 
(due to the coronagraph mask), 
the present data provide the first 
imaging polarimetry of the $\beta$ Pic disk at $r = 50$--120 AU with 
a resolution of $\sim$4 AU, covering the possibly 
interesting transition zone around 100 AU.

\section{OBSERVATIONS AND DATA REDUCTION}
The {\sl K\,} band polarimetric imaging of $\beta$~Pic was carried
out on 2003 January 4, using the CIAO mounted on the Subaru Telescope 
(Tamura et al. 2000). 
The AO system was utilized under the 
natural seeing of $0\farcs9$ at optical wavelengths, 
and 
the FWHM of about
$0\farcs19$ (= 3.7~AU) was achieved in
the {\sl K\,} band in spite of the large airmass (3.1--3.7) of $\beta$~Pic. 
The pixel scale was $21.3$ mas  pixel$^{-1}$ on the 
1024$\times$1024 InSb 
ALADDIN~\newcounter{two}\setcounter{two}{2}\Roman{two} array.
We employed an occulting mask of $0\farcs8$ diameter. The mask was not 
completely opaque and had a transmission of $\sim$2\%, which enabled us to
measure the stellar position accurately. We used a circular Lyot stop 
that blocked out the outer 20\% of the pupil diameter. 

The polarimeter consisted of a warm halfwave plate and a cold 
wire grid analyzer
(Tamura et al. 2003). 
The instrumental polarization was below 1\%, as measured 
during the commissioning of the polarimeter. 
The exposures were performed at 
four position angles (P.A.'s) of the halfwave plate, in the sequence  
of P.A.~= ~0$\arcdeg$, 45$\arcdeg$, 22$\arcdeg$.5, and 
67$\arcdeg$.5 to measure the Stokes $I$, $Q$, and $U$ parameters. 
The integration time for each frame was 10~s (1~s $\times$ 10 coadds),
and six frames were taken before rotation of the modulator. The integration
time per modulator cycle was 240~s, and six cycles of data
were collected.
After the imaging of $\beta$~Pic, HR 1998 (A2 V) was observed as a 
point-spread function (PSF) 
reference star. 
The integration times for
each frame and for each   
modulator cycle were the same as those for $\beta$~Pic, 
and four cycles of data were collected.

The obtained frames were calibrated in the standard manner using
\verb|IRAF|; dark subtraction, flat-fielding by dome flats, bad pixel
substitution, and sky subtraction. 
The sky brightness was measured as the constant mode value in the 
region where no emission was found in the field of view.
There was no significant change of the sky during obtaining images
in each cycle.
The Stokes parameters $(I,\, Q,\, U)$, the degree of polarization $p$, and 
the polarization angle $\theta_p$ were calculated as follows:  
\[ Q = I_{0} - I_{45}, \ \  U = I_{22.5} - I_{67.5},\] 
\[ I = (I_{0} + I_{45} + I_{22.5} + I_{67.5} ) / 2,\] 
\[ I_{\rm disk} = I_{\rm disk + PSF} - I_{\rm PSF},\] 
\[ p = \sqrt{Q^2 + U^2}/I_{\rm disk}, {\rm ~and}\] 
\[ \theta_p = (1/2)\, \arctan (U/Q) \ \ .
\]
The most dominant is the bright stellar halo, and the calculation of $Q$
or $U$ values corresponds to the subtraction of the
unpolarized stellar PSF. Because of a temporal variation in the PSF, 
we rejected data that had a large intensity ratio between the 
central peak and the halo at $r = 1\farcs5$ compared with that of
other data, and we checked the
resultant $Q$ and $U$ images. 
We performed frame registrations using
the \verb|IRAF/RADPROF| tool by measuring the position of the star inside
the mask. We then combined images for each modulator P.A., and 
calculated the $Q$ and $U$ images for each cycle. 
The resultant images were binned by 9 pixels ($\sim 0\farcs19$) with the 
\verb|IRAF/BLKAVG| tool. Finally, the $Q$ and $U$ images were averaged 
over all the cycles.

The Stokes $I$ parameter of the disk was estimated after the subtraction of the 
unpolarized halo of the central star. The reference PSF was obtained 
from the $I$ image of HR~1998. The images of HR~1998 were registered, 
rotated so as to adjust the P.A.'s of the spider patterns with each other, and   
then combined. 
The PSF was subtracted after the stellar position, spider pattern
direction, 
and the halo flux at $r = 1\farcs5$ of the PSF were matched
to those of the 
$I$ of $\beta$~Pic in each modulator cycle.
We checked each PSF-subtracted image and adjusted the scaling of the PSF 
by eye so that the halo of $\beta$~Pic was satisfactory subtracted.
The PSF-subtracted images 
of $\beta$~Pic were binned by 9 pixels and the images of all modulator
cycles were combined into one image.  
The total integration time was then 16.3 minutes for $\beta$ Pic and
9.8 minutes for the PSF reference star.  
The reference PSF was elongated along a direction different from that of 
$\beta$~Pic, possibly because these stars were observed at 
different elevations. In addition, the PSF mismatch between $I_0$ and
$I_{45}$, or between $I_{22.5}$ and $I_{67.5}$, caused residuals around the masked
region. Therefore, the inner region $r \leq 2\farcs6 $ cannot be 
discussed. A ghost image produced by the wire grid appears at 
$\sim$1$\farcs$1 from the star, but does not affect our results as  it
is located at $r < 2\farcs6$.  
Considering the uncertainty caused
by the PSF-subtraction (the registration, the scaling of the
PSF, and the mismatch of the PSF shape), we estimated the
uncertainty of $I_{\rm disk}$ as to be $\sim$20\%.

The polarization degree $p$ was corrected with 
the efficiency of 97\%  
in the {\sl K\,} band, which was measured during the commissioning of the polarimeter. The
interstellar polarization is negligible because of the proximity of  
$\beta$~Pic, $\sim0.01$\% at optical wavelengths (Krivova et al. 2000). 
We calculated uncertainties in the degree of polarization
by considering  
the standard deviations of the averages of $Q$ and $U$ over six cycles, 
and the uncertainty caused by the PSF subtraction for $I_{\rm disk}$.
Although we cannot estimate the consistency of each cycle
because of its low signal-to-noise ratio, we confirmed
the robustness of our results by calculating the polarization
using only any of four cycles.
%

\section{RESULTS}
\subsection{Intensity of Circumstellar Disk} 
We present the intensity image of the edge-on disk around 
$\beta$~Pic in Figure~1. The scattered light arising  
from the dusty disk is detected from $r=2\farcs6$ (50~AU) out to 
$r\sim6\farcs3$ (121~AU). The outer radius is determined on the basis of the observed 
sensitivity, as is clear from the previous optical observations showing
the outer disk of $r > 800$~AU 
(e.g., Smith \& Terrile 1984). 

Our adaptive optics observations detected the inner disk, $50 < r <
120$~AU, with a resolution of $\sim3.7$~AU. Three observations 
have been reported at optical and near-infrared wavelengths, which are comparable
to our results in terms of the observed inner disk and the spatial resolution
(Heap et al. 2000; Mouillet et al. 1997a, 1997b).
Our {\sl K\,} band image of the inner disk is substantially consistent
with the reported images.
The ``warp structure'' seen  in these images is also confirmed 
in the {\sl K\,} band.  

We estimated the P.A. of the disk midplane
by a least squares fitting under the
assumption that the disk has its maximum intensity along the 
midplane (e.g., Kalas \& Jewitt 1995). The fitting was performed on the
unbinned $I$ image, and on the northeastern and southwestern
wings at the same time. 
First, we assumed the P.A. of the midplane roughly, and we detected
the pixel position on the detector if the pixel showed
a maximum value in each radial step of $0\farcs05$ over the disk
vertical range of 1$\arcsec$.  The linear fitting to the spatial
distribution of such pixels gave the more precise PA of $30\fdg4
\pm 0\fdg2$ through the central star, measured for $r \leq 6\farcs0$. 
This value is consistent with the previous
optical and near-infrared estimates for the inner disk.
We adopted $30\fdg4$ as the P.A. of the disk midplane 
for a comparison with a
theoretical model, as described later. 

Figure~2 shows the surface brightness profile 
measured along the disk midplane. 
The slope of the profile is measured by a least squares fitting in the 
radial range of $2\farcs8 \leq r \leq 6\farcs0$, as
$r^{-1.86\pm0.04}$  and $r^{-1.68\pm0.07}$ in the northeastern and 
southwestern wings,
respectively.
These slope values are in good agreement with those at optical wavelengths
(${-1.79\pm0.01}$ and ${-1.74\pm0.04}$ in the northeastern 
and southwestern wings, respectively,
at $2\farcs8 \leq r \leq 6\farcs0$).
The {\sl K\,}$'$ band slopes reported by Mouillet et al. (1997a) are
fitted in different radial ranges, and, thus, cannot be directly
compared with ours.

Note that the profile reveals a clear difference from that seen 
in the younger disks around pre-main-sequence 
stars (e.g., Augereau et al. 2001); 
the profile appears to have ripples rather than a smooth decline.
We discuss the origin of the ripples in \S 4.

The intensity levels in the northeastern and 
southwestern wings are similar at equal
radii for $r \leq 6\farcs0$.
This is substantially consistent with the $HST$ result.
We could not confirm the large ($\sim$0.5 mag) intensity asymmetry
seen in the {\sl K\,}$'$ band  intensity profile reported by Mouillet et al. (1997a).
 
The vertical thickness (FWHM) of the edge-on disk was measured by
fitting a Gaussian function. The fitting was applied to each radial bin
of 9 pixels within a radius 80~AU where the signal-to-noise ratios
are sufficiently high. 
Since we could not find any dependence of the FWHM
on the radius, the obtained FWHMs were averaged, resulting in
18 and 19~AU in the northeastern and the southwestern disks, 
respectively. 
These FWHM values are consistent with the previous
optical (15--18 AU) and near-infrared (20 AU) data.

\subsection{Polarization of Circumstellar Disk}
The {\sl K\,} band polarization map of $\beta$~Pic is presented in 
Figure~3. The observed emission is polarized nearly
perpendicular to the radial direction, indicating that grains
in the disk scatter the light from the central star.
Although there are some regions where vectors deviate
from the centro-symmetric pattern, they are confined to regions 
where the signal-to-noise ratios are relatively poor.
Thus, the $\beta$~Pic disk is seen as an infrared reflection nebula
as in the case at optical wavelengths 
(Gledhill et al. 1991; Wolstencroft et al. 1995).
 
We show the {\sl K\,} band degree of polarization along the disk midplane 
in Figure~4. We detected about 10\% polarization in $50 
< r < 120$~AU. 
The degrees of polarization are roughly the same for
the northeastern and southwestern wings, 
as for the radial surface brightness levels.
In contrast, the optical polarization data
show a substantially higher level of polarization
($\sim$15\% at $r > 150$~AU) and
a higher polarization with an asymmetrical distribution 
in the southwestern wing.  
Since the uncertainty is large in the outer southwestern wing
($r > 100$~AU),
we cannot claim that the polarization increases with radius.
Note that there is a dip in the polarization levels around
$r \sim$ 100 AU in both wings.

\subsection{Model}
In this section, we compare the results of our polarization measurements with 
detailed model calculations 
to gain an insight into the properties of dust grains in the debris disk 
around $\beta$ Pic.
We code for model polarized intensities by incorporating Mie theory 
into the algorithm that is described in Artymowicz et al. (1989). 
Numerical calculations require reasonable assumptions concerning the spatial 
and size distribution of dust grains (Artymowicz et al. 1989; Artymowicz 1997). 
Krivova et al. (2000) have successfully reproduced the dependences of 
linear polarization on wavelength and radial distance by assuming the presence 
of homogeneous spherical silicate grains throughout the disk. 
We calculated the Stokes parameters $(I, Q, U)$ using Mie theory and 
subsequently the degree of linear polarization $P=\sqrt{Q^2+U^2}/I$
by employing their model parameters : 
The disk is inclined by $4^\circ$ with respect to the line of sight 
and truncated at $r=1000$~AU in cylindrical coordinates $(r,z)$, 
the radius of dust grains ranges from ${a}_{\min}=5$~nm 
to ${a}_{\max}=100~\mu$m, 
and the spatial and size distribution $n\left({r,z,a}\right)da$ 
in the range of dust radius from $a$ to $a+da$ is described 
as 
\begin{eqnarray} n\left({r,z,a}\right) & = & {n}_{0} 
\exp \left\{{-\left[{\frac{|z|/{r}_{0}}{0.05\left(r/{r}_{0}\right)^{\eta}}}\right]^{1.1}}\right\} 
\left[\left(\frac{r}{{r}_{0}}\right)^{-1}+\left(\frac{r}{{r}_{0}}\right)^{2.7}\right]^{-1} {a}^{-3.5}, 
\label{n} 
\end{eqnarray}
where ${r}_{0}=116$~AU (i.e., 6\arcsec) 
and $\eta=1.2$, as used for the outer disk model of 
Krivova et al. (2000). 
In the inner disk at distances of $r \le {r}_{0}$, however, 
we set $\eta=0$, as inferred from our near-infrared measurements 
of the inner disk. 
Stellar radiation pressure reduces the number density of grains 
with $a\le a_0$ by blowing them out of the system. 
Here $a_0$ is the dust radius at which the ratio of radiation pressure 
to stellar gravity reaches 0.5 (see, e.g., Zook \& Berg 1975). 
Note that once the composition of grains is assumed, 
the ``blown-out'' radius $a_0$ and the amount of the reduction 
in the number density are no longer free parameters. 
We take the refractive indices of ``astronomical silicate'' 
from Laor \& Draine (1993), for which $a_0=2.0~\mu$m 
(see Kimura \& Mann 1999). 
An estimate of the reduction factor for the density ${n}_{0}$ 
is not trivial, because a nonlinear equation along with assumptions 
for unknown sources of dust has to be solved (Krivov et al. 2000). 
Here we simply reduce the number density for dust grains having 
radius $a \le a_0$ by a factor of 10 in the northeastern wing of the disk 
and by 15 in the southwestern wing.  
We note that our observed $K$ band polarization turned out to be 
inconsistent with the reduction factors of approximately 15 
(northeastern) 
and 20 (southwestern) adopted by Krivova et al. (2000).

Figure~5 shows the model polarization in the midplane 
($solid curves$) plotted together with our near-infrared data 
($open squares with error bars$) as well as currently available 
optical data from Gledhill et al. 
(1991; $filled circles with error bars$) and 
Wolstencroft et al. 
(1995, open circles with error bars). 
Prior to calculations of the polarization, the derived Stokes 
parameters were integrated over $1\arcsec$ 
both optical and infrared data
in a direction 
perpendicular to the midplane. 
Note that the model properly reproduces 
the absolute values and spectral variation of the linear 
polarization, as well as its 
northeast-southwest asymmetry at optical wavelengths and 
northeast-southwest symmetry at 
near-infrared wavelengths.
We also found that the radial profile of the observed {\sl K\,} band 
intensity agrees with our numerical results for the Stokes $I$
parameter (Figure~6). 
Our numerical simulations in the optical wavelength range 
provide a neutral color in the outer disk, consistent 
with the multiple-wavelength data from 
Paresce \& Burrows (1987) and Lecavelier des Etangs et al. (1993). 
The color becomes slightly redder in the inner part of the disk, 
but we cannot reconcile the strong decrease in the {\sl B\,} band 
brightness that was observed by Lecavelier des Etangs et al. (1993).  

\section{DISCUSSION}
The observed {\sl K\,} band polarization is characterized by its low 
value and northeast-southwest symmetry, 
in contrast to the high value and asymmetrical distribution of the 
optical polarization.
To better understand this wavelength dependence, 
we consider the blown-out radius in comparison with the radius of dust particles that 
mainly determines the degree of linear polarization at each wavelength 
of the observations.
It is worth noting that dust grains in the Rayleigh scattering 
regime ($a \ll \lambda$) as well as in the geometrical optics regime 
($a \gg \lambda$) produce a polarization much higher than that observed, 
while grains in the Mie scattering regime ($a \approx \lambda$) lower 
the polarization (Voshchinnikov \& Kr\"{u}gel 1999).
At optical wavelengths, Mie scattering particles are smaller than the 
blown-out radius (i.e., $a < a_0$) and therefore their abundances are 
lowered.
As a result, the optical polarization is kept high, in particular, in 
the southwestern wing, where the number of Mie scattering particles is greatly 
reduced.
Provided that the radius of Mie scattering particles is comparable to 
or larger than the blown-out radius (i.e., $a \ga a_0$), the 
polarization becomes low and symmetrically distributed.
The low and northeast-southwest 
symmetrically distributed polarization observed in the {\sl K\,} band 
implies that the blown-out radius $a_0$ lies in the range of 
microns, consistent with $a_0=2~\mu$m in our model.

The model uses compact, spherical, optically bright 
silicate grains whose properties are in accord with 
the high albedo estimated by Backman et al. (1992). 
Optically bright compact grains are consistent with 
the blown-out radius of a few microns necessary 
for explaining the observed low {\sl K\,} band polarization. 
Regardless of our successful results, however, 
spherical silicate grains are not necessarily representative 
of real dust particles in the $\beta$ Pic debris disk. 
Mid-infrared spectroscopic observations have revealed a 
similarity between dust in the $\beta$ Pic debris disk 
and cometary dust in the solar system (Knacke et al. 1993). 
Cometary dust is likely an optically dark fluffy aggregate consisting 
of submicron-sized grains (Greenberg \& Gustafson 1981;
Kimura et al. 2003).
In a model with dust aggregates for the $\beta$ Pic system, 
a problem is encountered in that the blown-out radius of fluffy 
aggregates is much larger than a few microns 
(cf. Kimura \& Mann 1999). 
To resolve this apparent discrepancy, we here consider 
icy parent bodies ``planetesimals'' that consist of ice 
and dust, analogous to cometary nuclei in the solar 
system (see Yamamoto 1985). 
After ejection from their parent bodies, dust particles 
are still encased in ice at the distances of our observations, 
although we have to keep in mind the possibility that 
the intense ultraviolet radiation from $\beta$ Pic
erodes icy grains by photo-sputtering (see Artymowicz 1994). 
If ice fills the pores of optically dark dust aggregates, 
then the dust may resemble optically bright compact particles. 
The refractive indices of ice with optically dark inclusions 
are close to those of silicates, depending on the volume 
fraction of the inclusions (see Mukai et al. 1986). 
As a result, the blown-out radius of ice-filled, optically 
dark aggregates would lie in the range of microns, 
as expected for dirty water-ice spheres (cf. Artymowicz 1988). 
Our results may indicate that compact spherical silicate 
grains are simply a good approximation for ice-filled 
fluffy aggregates of optically dark submicron grains 
existing in the $\beta$ Pic system.  

Although the smooth model curves fit the intensity and 
linear polarization profiles in the {\sl K\,} band well, 
a noticeable dip appears in both the intensity and 
polarization profiles at $5\farcs2$--$5\farcs4$ (i.e., 102--104~AU) 
along both directions (Figs. 2 and 4), except for
in the SW intensity profile. 
The dip is also seen in the PI profile, indicating that
it is not caused by an artifact due to the PSF subtraction
for $I_{\rm disk}$.
We here interpret the dip in terms of a gap in the radial 
density distribution of dust grains around 100 AU from 
the central star. 
Keeping in mind that the observed radiation is an integrated 
quantity over a line of sight, the effect of a gap on the 
integrated quantity depends on its integrand. 
By analogy to interplanetary dust or cometary dust, we expect 
the integrand of intensity to show a strong forward scattering 
enhancement and the integrand of polarization to peak 
around a scattering angle of $90^\circ$  (Artymowicz 1997). 
Although a deficit of dust grains reduces the intensity, 
forward scatterers residing far from the gap region tend 
to obscure the reduction. 
This may explain why the dip in the intensity 
profile is seen only on the northeast side of the disk where the 
signal-to-noise ratio is higher than the southwest side. 
When the line of sight is tangent to a gap in the radial dust density, 
a low degree of polarization results from a lack of dust grains 
having scattering angles around $90^\circ$.
On the other hand, an erroneous hump in the intensity profile
would decrease the polarization, but simultaneous decreases 
in both the intensity and the polarization do not seem to be coincidental. 
Mouillet et al. (1997a) measured the {\sl K\,} band surface 
brightness of the $\beta$ Pic disk, in which we confirm 
the presence of a similar dip around $5\farcs2$. 
It is highly unlikely that both we and Mouillet et al. (1997a) 
erroneously obtained a dip at the same distance 
from the star on the basis of independent data and analyses. 
Therefore, this dip is not an artifact arising from our data 
analysis, but is a real signature corresponding to the depletion of 
dust grains around 100~AU. 
We expect that dust grains are located near their parent-body 
planetesimals, because grains are destroyed by mutual 
collisions before being transported inward by the Poynting-Robertson 
effect in the $\beta$ Pic system (Artymowicz 1997; Krivov et al. 2000). 
Consequently, the dip in the intensity and polarization profiles 
around 100~AU is evidence for a paucity of planetesimals 
in this region.  

Our {\sl K\,} band intensity data reveal 
not only the dip around 100~AU but also ripple structures 
over the entire observed disk region between 
50 and 121~AU (see Fig. 2). 
Comparing our intensity data with the results of Mouillet et al. (1997a), 
we realize that there is a great similarity between the ripple structures 
obtained in the two 
independent observations. 
If all the humps in the ripple structures are in fact planetesimal 
belts, then our data indicate that the $\beta$ Pic disk 
consists of multiple planetesimal belts. 
Several planetesimal belts (rings) have already been identified from 
mid-infrared observations of the $\beta$ Pic disk 
in the inner region ($r<90$~AU)
(Wahhaj et al. 2003; Okamoto et al. 2004; 
Telesco et al. 2005). 
Recent observations have revealed that the $\beta$ Pic debris disk 
resembles the circumstellar disk around AU Microscopii, 
including spatially localized enhancements and deficits 
(Liu 2004; Metchev et al. 2005). 
Therefore, we conclude that the presence of multiple planetesimal 
belts is common in circumstellar disks around Vega-type stars.

%
%
%
%
%
%

\acknowledgments

M. T., H. K., and T.Y. are supported by Grants-in-Aid from the Ministry
of Education, Culture, Sports, Science, and Technology.
M. F. is supported by the JSPS Research Fellowships for Young
Scientists.

\begin{figure}
\plotone{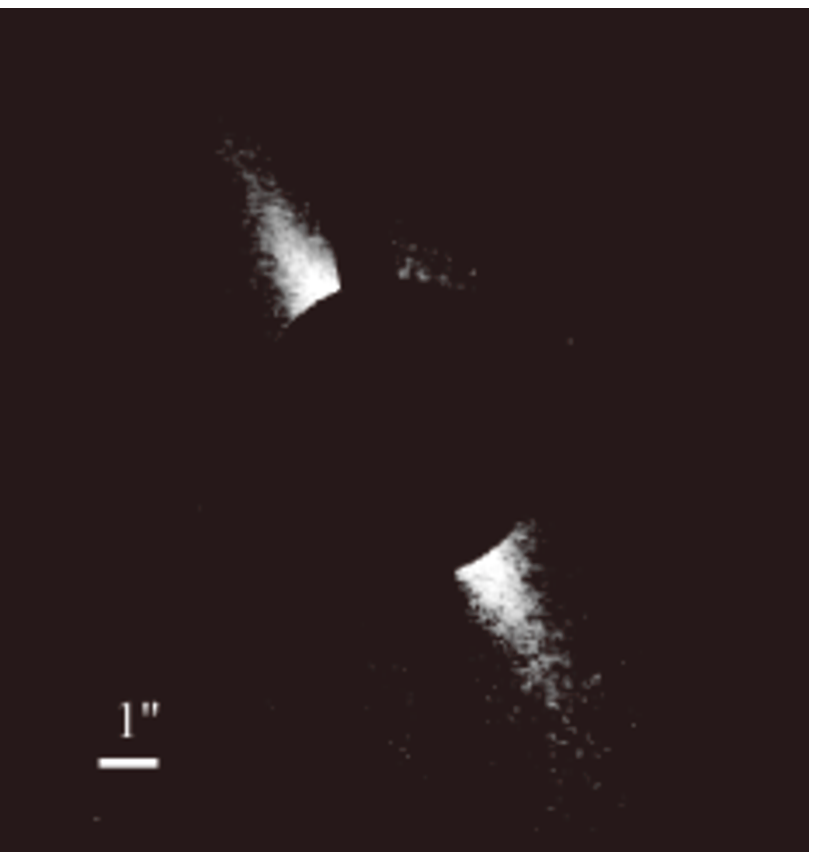}
\caption{{\sl K\,} band image of edge-on disk around
 $\beta$~Pic. The inner region ($r \leq 2\farcs6 = 50$~AU) was 
 software masked (see text).
The direction of the
 secondary spider pattern was also masked. The pixel scale is  
 21.3~mas~pixel$^{-1}$. The image is linearly scaled and displayed from
 $+3$ to $+15\sigma$. North is up and east is to the
 left. \label{fig:bpic_i}} 
\end{figure}

\clearpage

\begin{figure}
\plottwo{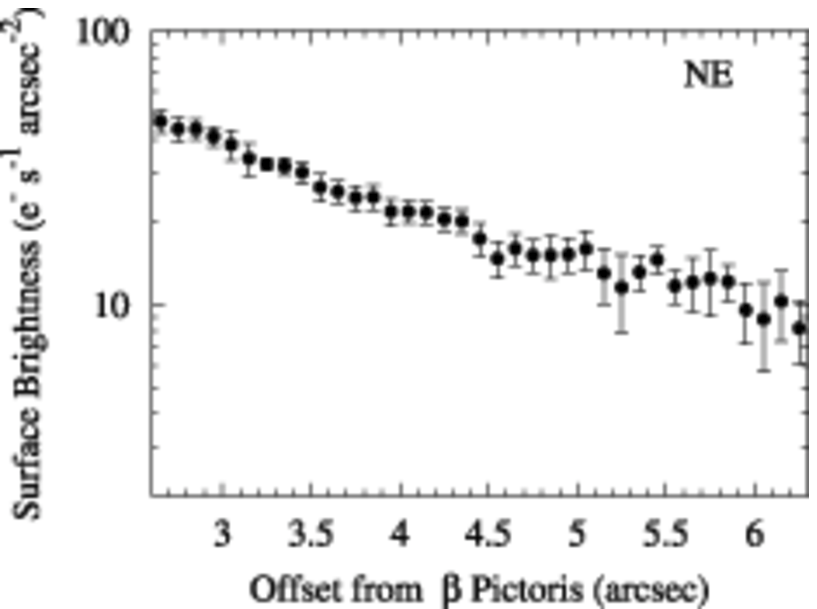}{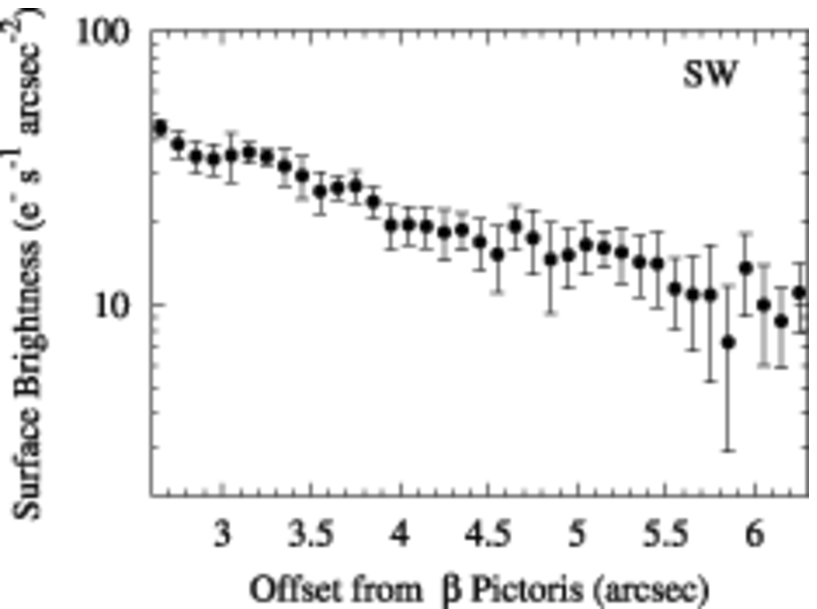}
\caption{Surface brightness distribution along disk midplane for
 northeastern disk ($left$) and southwestern disk ($right$).  The
 surface brightness was averaged over the disk vertical  thickness  of
 $0\farcs19$ and the radial range of  $0\farcs1$. \label{fig:rprof_i}}  
\end{figure}

\clearpage

\begin{figure}
\plotone{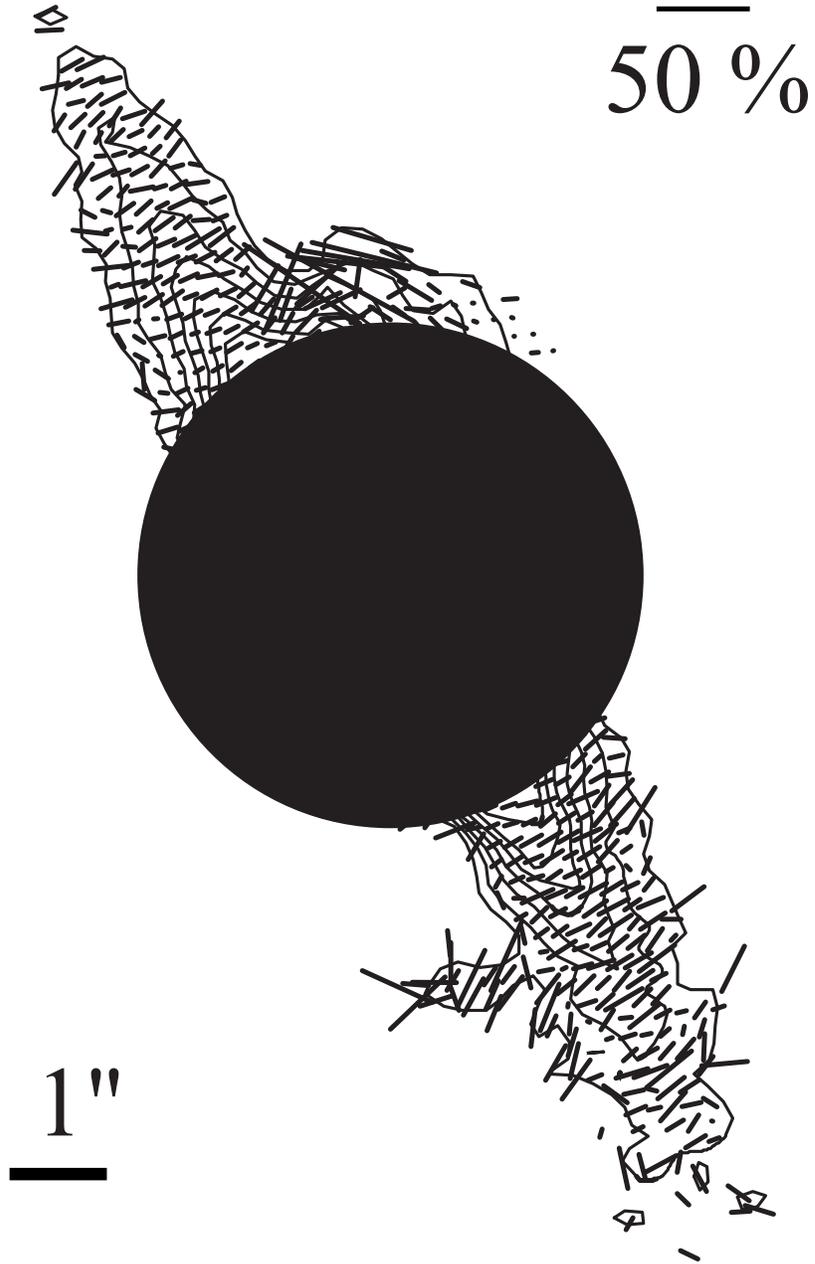} 
\caption{Polarization vector map in {\sl
 K\,} band, overlaid with intensity contours. The polarization was
 calculated using the binned images ($9  \times 9$ pixels), thus the
 spatial scale is $0\farcs19$. The intensity contours are created from 
 the Stokes $I$ parameter image with a raw pixel scale ($0\farcs0213$), and stars from 
 $3$ to $15\sigma$ with a spacing of $2\sigma$.  Note that the 
 direction of ${\rm P.A.}\sim 10\arcdeg$ is affected by the spider
 structure. 
\label{fig:bpic_p}}
\end{figure}

\clearpage

\begin{figure}
\plotone{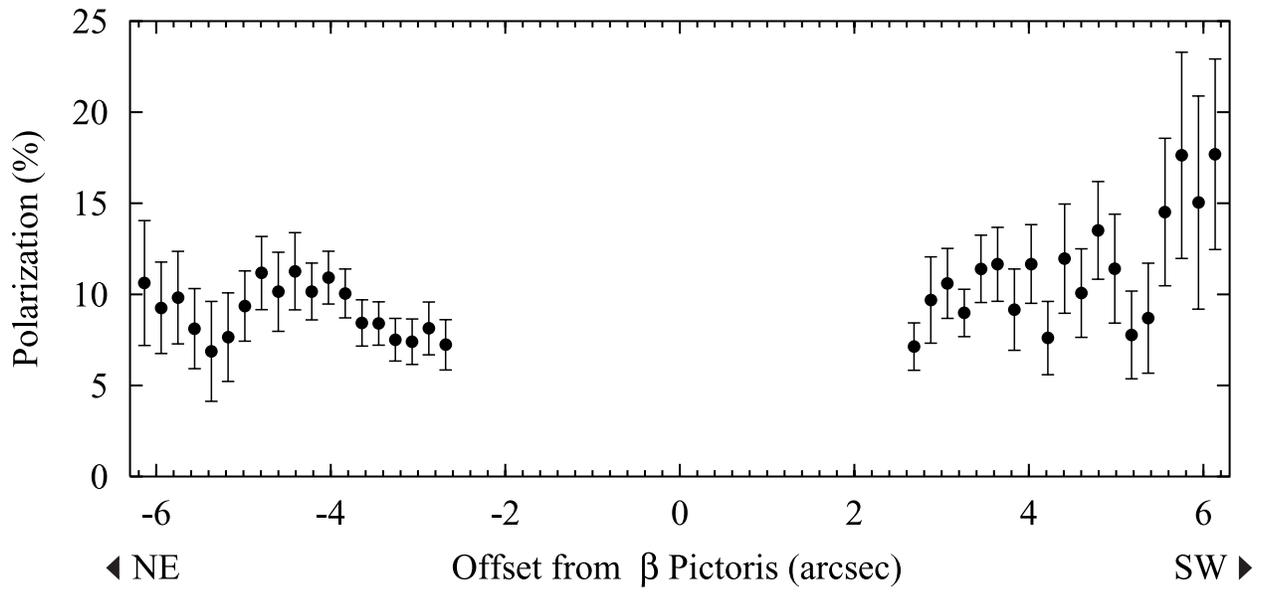}
\caption{Degree of polarization in {\sl K\,} band measured along 
 disk midplane. 
\label{fig:bpic_vec}}   
\end{figure}

\begin{figure}
\plotone{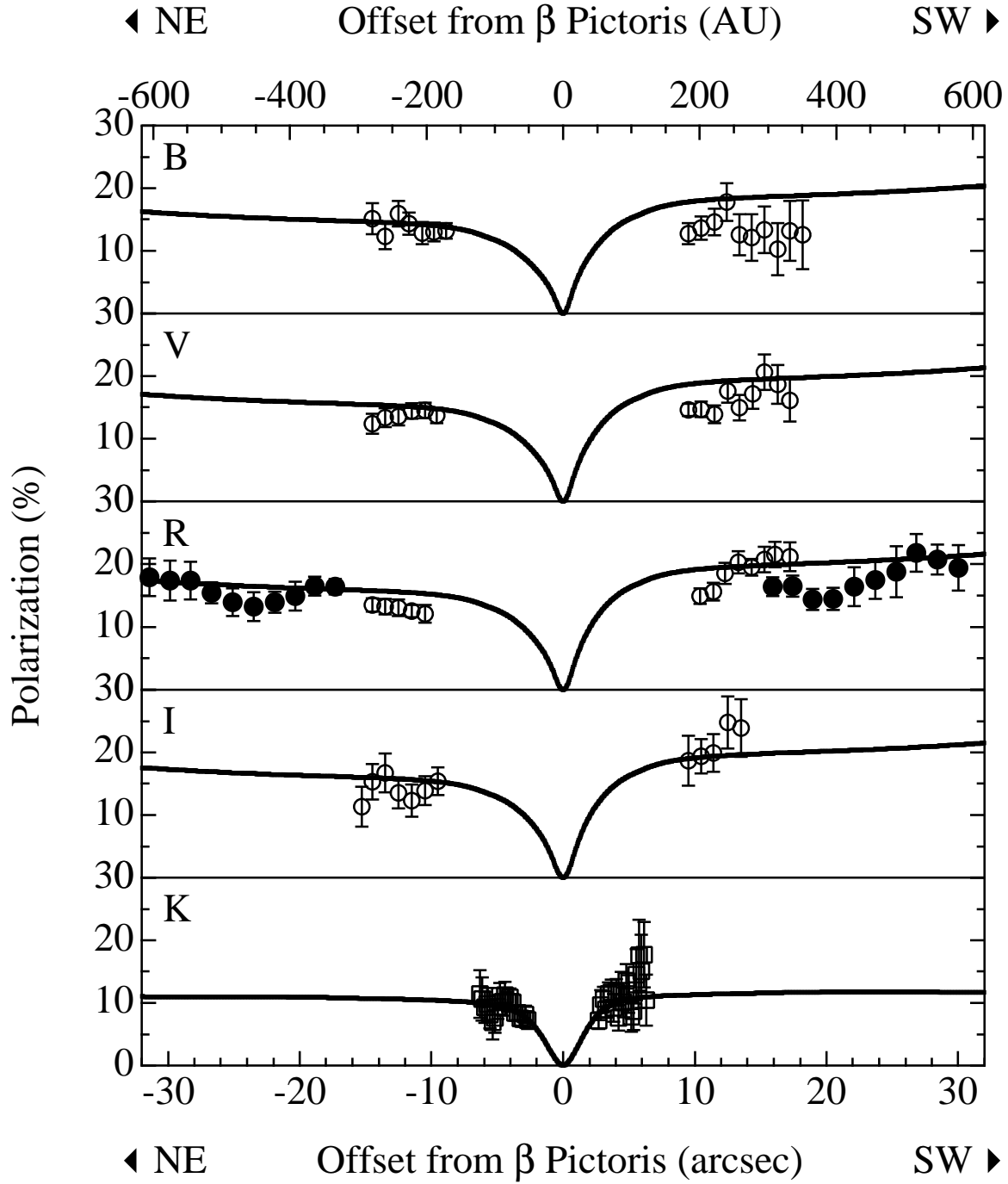}
\caption{
Linear polarization as a function of offset from $\beta$ Pic. 
{\it Solid curves}: Our model results; 
{\it squares}: our {\sl K\,} band data; 
{\it filled circles}: {\sl R\,} band data from Gledhill et al. (1991); 
{\it open circles}: {\sl B\,}, {\sl V\,}, {\sl R\,}, and {\sl I\,} band 
data from Wolstencroft et al. (1995).
\label{fig5}}
\end{figure}

\begin{figure}
\plotone{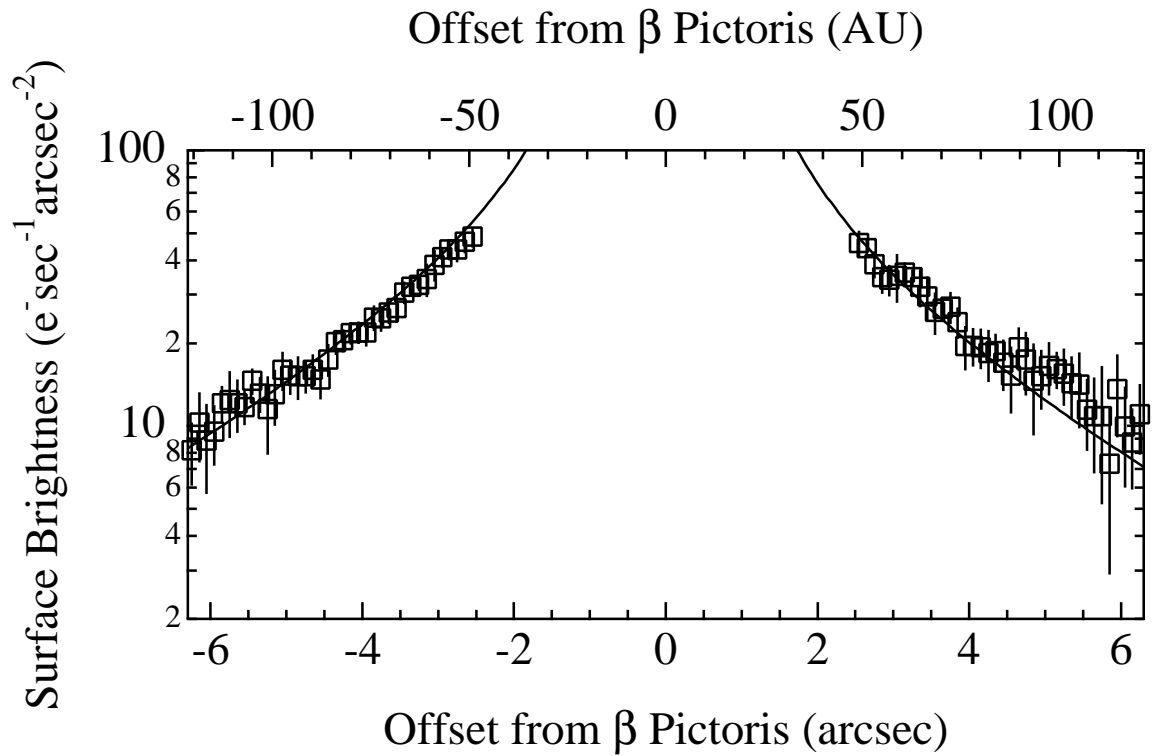}
\caption{
{\sl K\,} band intensity as a function of offset from $\beta$ Pic. 
{\it Solid curves}: Our model results;
{\it squares}: our {\sl K\,} band data. 
\label{fig6}}
\end{figure}

\end{document}